# Interparticle interaction and size effect in polymer coated magnetite nanoparticles

M. Thakur, K. De, and S. Giri
Department of Solid State Physics, Indian Association for the Cultivation of Science, Jadavpur, Kolkata 700 032, India
S. Si, A. Kotal, and T. K. Mandal
Polymer Science Unit, Indian Association for the Cultivation of Science, Jadavpur, Kolkata 700 032, India

**Abstract**
Magnetization and Mössbauer studies have been performed on the polymer coated magnetite nanoparticles with particle size from 5.1 to 14.7 nm. The maximum in the temperature dependence of magnetization ($T_M$) is found to be inconsistent with the particle size ($D_{TEM}$). The effective magnetic anisotropy ($K_{an}$) is found to increase with the decrease of $D_{TEM}$, which is attributed to the increase of surface anisotropy. The absence of coercivity and remanence of magnetization noticed well above $T_M$ indicates the superparamagnetic behaviour, which has also been observed in the temperature dependent Mössbauer results. The temperature dependence of hyperfine field is found to follow similar dependence of saturation magnetization for bulk magnetite.

## 1. Introduction

The investigations on nano-scale magnetism have been the subject of intense research because of the unusual magnetic properties, which are significantly different from their bulk properties [1]. In case of magnetic nanoparticles, the interplay between the intrinsic properties and interparticle interactions are crucial to control the overall magnetism. The intrinsic properties are strongly dependent on the size, shape, and nature of the single domain structure [2]. On the other hand, the recent studies have shown that the magnetic properties of nanoparticles are also extremely sensitive to the interparticle interactions [3-8].

The investigation on the magnetic properties of magnetite is one of the central issues in magnetism because of the technological importance as well as the interesting fundamental properties [9]. Even today, the low temperature properties of the compound is not well understood, which still remain at the focus of active research. Here, magnetite has been chosen for the study to investigate the role of particle size dependence and interparticle interaction in magnetic nanoparticles. Magnetite belongs to the spinel ferrite materials group, which exhibits cubic structure with space group $Fd\bar{3}m$ at room temperature [10]. The general chemical formula of magnetite is $Fe_3O_4$, written as $Fe^{3+}[Fe^{2+},Fe^{3+}]O_4$ from ionic point of view, where octahedral ions in the bracket occupy $16d$ (B) site and $8a$ (A) site is occupied by the tetrahedral ions before bracket. Since $Fe^{2+}$ and $Fe^{3+}$ coexist at the same cryatallographic site, the structure is called as inverse spinel structure. Magnetite orders ferrimagnetically below $T_N$ = 860 K, where tetrahedral and octahedral ions are aligned ferromagnetically within each sublattice and antiferromagnetically between the two sublattices [11,12]. The magnetic structure agrees with the saturation magnetic moment, ~4 $\mu_B$/ion, determined experimentally using neutron diffraction results [12,13]. Magnetite also undergoes another Verwey transition at $T_V$ ~ 120 K, where the charge ordering between $Fe^{2+}$ and $Fe^{3+}$ has been observed below $T_V$ [14]. However, the recent experimental results demonstrate the lack of ionic charge ordering [15].

In case of magnetite nanoparticles the magnetic properties display wide varieties of interesting properties in contrast to their bulk counterpart, where magnetism of those particles were strongly sensitive to the different synthesis procedures [16,17]. In the present study, the magnetism of the polymer coated magnetite nanoparticles have been investigated on the particles of average diameter ($D_{TEM}$) with 5.1, 7.0, 10.5, and 14.7 nm using magnetization and Mössbauer studies, where magnetic properties are discussed by focusing the role of particle size dependence and the nature of the interparticle interaction.

## 2. Experimental procedures

Nearly monodisperse magnetite nanoparticles of variable sizes were synthesized by alkaline hydrolysis of iron(II) ions in presence of two polyelectrolytes, viz., poly(acrylic acid) (PAA) and sodium salt of carboxymethyl cellulose (NaCMC) by the techniques described in our earlier report [18]. Single phase of the magnetite nanoparticle was characterized by X-ray powder diffraction and electron diffraction using Transmission Electron Microscope (TEM) described in our earlier report [18]. The characteristic features of the magnetite nanoparticles taken for the present study are given in Table 1. The samples have been defined as P5, P7, P15, and N10 for the simplicity, where P5, P7, P15, and N10 stand for the PAA coated magnetite particles of $D_{TEM}$ with 5.1 nm, 7.0 nm, 14.7 nm and NaCMC coated particle of $D_{TEM}$ with 10.5 nm, respectively. The attachment of the polymer on the particle surface was confirmed by FTIR spectroscopy [18]. The amount of the polymer content (given in Table 1) was estimated by thermogravimetric analysis using a Mettler Toledo Star System TGA/SDTA851e in presence of $N_2$ gas. The magnetization study was performed using a commercial SQUID magnetometer (Quantum Design, MPMS-5). The Mössbauer spectrum was recorded in a transmission geometry using a ~ 370 MBq $^{57}$Co source in a Rh matrix with a Wissel velocity drive unit in a constant acceleration mode. The Mössbauer parameters were estimated with respect to $\alpha$-Fe.

## 3. Experimental results

In order to understand the magnetic properties of the magnetite nanoparticles the magnetization was measured as a function of temperature and magnetic field. The field cooled (FC) effect of magnetization as a function of temperature is shown in Fig. 1 for P5, P7, P15, and N10 measured at 0.01 T under both zero-field cooled (ZFC) and FC conditions. In case of ZFC condition the samples were cooled down to the desired temperature without magnetic field and the magnetization were measured in the heating cycle after the application of magnetic field, while for the FC measurement the samples were cooled down to the desired temperature with magnetic field and the magnetization were measured in the heating cycle like a ZFC measurement. The onset temperature of branching between ZFC and FC magnetization ($M_{FC}$) is noticed above the broad maximum ($T_M$) in the temperature dependence of ZFC magnetization ($M_{ZFC}$) for P5, P7, and N10, while $T_M$ is not observed for P15 up to 300 K. The values of $T_M$ were estimated from the change of sign of $dM_{ZFC}/dT$ in the plot of $dM_{ZFC}/dT$ against temperature, which are given in Table 2. The values of $T_M$ were found to be 162 K, 142 K, and 154 K for P5, P7, and N10, respectively, which do not indicate any consistency with the particle size. If it is assumed that the blocking temperature, $T_B$ is close to $T_M$ in the present observation, $T_B$ can be related as

$$K_{an}V/(k_B T_B) = \ln(tf_0), \qquad (1)$$

for a random distribution of particles with a single magnetic domain [19]. Here, $V$, $K_{an}$, $t$, and $f_0$ are the average particle volume, the effective anisotropy energy density of the particle, time of measurements, frequency at 0 K limit, respectively. The value of $\ln(tf_0) \approx 25$, is typically used assuming $t = 100$ s for dc magnetization measurement and $f_0 = 10^9$ s$^{-1}$. The values of $K_{an}$ are estimated to be $3.1 \times 10^5$ J/m$^3$, $1.8 \times 10^5$ J/m$^3$, and $1.0 \times 10^5$ J/m$^3$ for P5, P7, and N10, respectively, where the values are found to be larger than that of the reported values for magnetite nanoparticles [17,20,21].

Hysteresis of magnetization was measured well above and below $T_M$. In Fig. 2

the magnetization curves measured well above $T_M$ for P5, P7, and N10 do not show the coercivity and remanance of magnetization. In absence of remanence and coercivity, the magnetic size ($D_M$) of the particles are estimated from the initial susceptibility, $\chi_i = (dM/dH)_{H\to 0}$, which mainly arises from the largest particles. The upper limit of $D_M$ may be estimated using the formula,

$$D_M = [(18k_B T/\pi)(\chi_i/\rho M_s^2)]^{1/3}, \quad (2)$$

where $\rho$ is the density of $Fe_3O_4$ (5.18 g/cm$^3$). The values of $\chi_i$ were determined from the linearity of the magnetization curve near $H = 0$. Using the values of saturation of magnetization the estimated upper limit of $D_M$ are estimated to be 8 nm, 9, nm, and 10.5 nm for P5, P7, and N10, respectively, which are consistent with the average particle size obtained from TEM observation. The upper limit of $D_M$ for P5 and P7 are found to be slightly larger than those of $D_{TEM}$ with ~5.1 and ~7.0 nm. On the other hand, the value of $D_M$ is ~10.5 nm, which is equal to the value of $D_{TEM}$ with 10.5 nm for N10. In case of nanoparticle with distribution of particle size, the maximum magnetic size is determined by the larger particles of the distribution. Thus, slightly larger values of $D_M$ than the average value of physical size are reasonable for P5 and P7. On the other hand, the equal values of $D_M$ and $D_{TEM}$ noticed for N10 may indicate the formation of a surface shell with spin disorder, which may create a dead magnetic layer originated by the demagnetization of the surface spins.

In order to understand the nature of field dependence of magnetization in the superparamagnetic regime, the magnetization curves were analyzed using the law of approach to the saturation of an assembly of particles with uniaxial anisotropy [22],

$$M(H) = M_s(T)[1 - k_B T/M_s(T)vH - 4K_{an}^2/15M_s(T)^2 H^2] + \chi_0 H, \quad (3)$$

where $M_s(T)$ is the saturation magnetization at a particular temperature and $\chi_0$ is the high field susceptibility. The best fit of the magnetization curves using Eq. (3) are shown in Fig. 3 for P5, P7, and N10 by considering volume ($v$), $K_{an}$, $M_s(T)$, and $\chi_0$ as free parameters. The values of the magnetic size obtained from the fitting are almost same with those values obtained using Eq. (2), as seen in Table 2. The estimated values of $K_{an}$ seen in Table 2 are close to those found in the literature for magnetite nanoparticles [17,20,21]. The value of $K_{an}$ includes several intrinsic factors, which are mainly volume, surface, shape, and magneto crystalline anisotropies in case of noninteracting particles with a single magnetic domain. The effective anisotropy energy may be simplified as

$$E_{an} = K_v v + K_s s, \quad (4)$$

by neglecting the other terms, where $K_v$ and $K_s$ are the uniaxial volume and surface anisotropy constants, and $v$ and $s$ are the volume and surface area of the particle, respectively. For particles with average diameter $D$, the Eq. (4) gives

$$K_{an} = K_v + \frac{a_s}{D} K_s. \qquad (5)$$

By using Eq. (5) the values of $K_{an}$ obtained from Eq. (3) are plotted against the inverse of $D_{TEM}$ seen in Fig. 4, considering $a_s = 6$ for spherical particle, which holds satisfactorily for different nanomatirc particles [23]. The linear fit of the plot gives $K_v = 0.20 \times 10^3$ J/m$^3$ and $K_s = 0.15 \times 10^5$ J/m$^2$. The value of $K_v$ is found to be much smaller than the value of $K_{an}$ for bulk magnetite (~$0.9 \times 10^4$ J/m$^3$ at 300 K) [24]. If the particles are assumed to have perfect spherical shape, the symmetry arguments show that the surface anisotropy normal to the surface is zero. Therefore, the value of $a_s$ should not be 6 to get nonzero value of $K_s$, which may be a more complex function reflecting the particular magnetization reversal process due to the strong uniaxial anisotropy of surface atoms. In addition, the incoherent rotation of the spins on the surface of the particles may also lead to a more hysteretic behaviour resulting an additional contribution to the effective anisotropy. Thus, the simplified picture of Eq. (5) does not hold in the present observation in order to give realistic value of $K_v$.

The hysteresis of magnetization under ZFC condition is shown in Fig. 5 with small coercivity and remanence of magnetization at 5 K. The values of coercive field ($H_c$) and remanence of magnetization ($M_r$) are given in Table 2. The values of $H_c$ increase, while $M_r$ decrease with the decrease of $D_{TEM}$. The strong effective anisotropy enhances the coercivity [25], which is noticed in the present observation, where the coercivity increases with the decrease of particle size. The inset of the figure shows the hysteresis of magnetization with field up to 5 T, indicating a saturating tendency of magnetization around 5 T. However, the values of magnetization ($M_s$) at 5 K and 5 T for all the cases are much smaller than the saturation moment (~4 $\mu_B$) of bulk magnetite. As seen in Table 2, the values of $M_s$ increase with $D_{TEM}$ for the PAA quoted particles. On the other hand, the value of $M_s$ for NaCMC coated sample of $D_{TEM}$ with 10.5 nm is less than that of those PAA coated samples, which is even less than the value of P5 of $D_{TEM}$ with 5.1 nm.

In order to gain further insight into the magnetic properties, the Mössbauer study has been performed on the sample N10. Mössbauer spectra were measured under zero magnetic field in the temperature range between 4.2 K and 300 K. Different characteristics of the spectra with temperature are shown in Fig. 6. If the magnetic anisotropy energy is close to the thermal excitation energy, the fast fluctuation of the moment of fine particles with single magnetic domain exhibits superparamagnetic relaxation, which is expected to follow the Néel-Brown expression [26,27],

$$\tau(T) = \tau_0 \exp(E_{an}/k_B T), \quad (6)$$

where, $\tau$ is the superparamagnetic relaxation time and $E_{an}$ is the magnetic anisotropy energy. The value of $\tau_0$ is typically found to be in the order of $10^{-9} - 10^{-11}$ s, which is almost independent of temperature. In such a case, the observed magnetic features depend on the observation time of the experiment ($\tau_{obs}$). In case of Mössbauer study, the magnetically separated sextet pattern, bearing the characteristics of ordered state is observed, when $\tau \gg \tau_{obs}$ at a particular temperature. The features of superparamagnetic relaxation start to appear by reducing the splitting of the sextet pattern, when both the time scales become comparable. The sextet pattern collapses in to a quadrupole doublet or singlet at a particular temperature, while $\tau \ll \tau_{obs}$ in the extreme case. The appearance of mainly quadrupole doublet in Fig. 6 indicates the signature of superparamagnetic relaxation at 300 K. With the decrease of temperature the sextet pattern starts to develop associated with the decrease of line width. In case of Mössbauer study, the blocking temperature is obtained, when the equal area of the quadrupole doublet and magnetically separated components are found at a particular temperature [28]. Here, the blocking temperature observed from Mössbauer spectra is above 192 K, which is much larger than that of the maximum in the temperature dependence of magnetization ($T_M = 154$ K). As seen in Eq. (1) the observation of blocking temperature strongly depends on the time window of the experiment ($t$), where the value of $t$ for Mössbauer study ($\sim 10^{-6}$ s) is much less than the value of $t$ for magnetization study. Thus, the blocking temperature obtained from Mössbauer spectra is found to be larger than that of the value obtained from magnetization studies [29].

As seen in Fig. 7(a) the spectrum at 4.2 could not be fitted well by considering unique hyperfine field. In order to fit the spectrum satisfactorily, we fit the spectrum [Fig. 7(b)] using the least square fitting program NORMOS [30], where a distribution function, $P(H)$ of the hyperfine field has been taken into account. The nature of distribution of hyperfine field is shown in Fig. 8, where the range of the hyperfine field is observed in between ~45 T to ~55 T for the majority of the particles. However, the value of the average hyperfine field of the majority of the particles was close to the estimation of hyperfine field, $H_{hyp} \sim 51.71$ T from least square fitting by assuming unique hyperfine field, which is also close to the values for bulk magnetite [31]. The values of isomer shift (IS) and quadrupole splitting (QS) obtained from the least square fitting were given in Table 3 by considering unique hyperfine field. The values of IS were found to vary in the range 0.39 – 0.45 mm/s, while the value of QS is found in the range -0.02 – -0.03 mm/s at different temperatures as seen in Table 3.

The low energy collective excitations of an ordered magnetic system are known as spin waves or magnons. In the thermodynamic equilibrium, spin waves result in a decrease of the spontaneous magnetization with increasing temperature, which has the following form in the low temperature range [32,33],

$$M_s(T) = M_s(0)[1-BT^\epsilon], \qquad (7)$$

where $M_s(0)$ is the spontaneous magnetization at 0 K and B is a constant, which is closely related to the exchange intrgral, $J$ (B ~ $1/J^\epsilon$). Eq. (7) is known as Bloch $T^{3/2}$ law for $\epsilon = 3/2$, which has been verified experimentally for most of the bulk materials [32]. However, different values of $\epsilon$ have also been reported for some of the bulk spinel ferrites [32]. For fine particles and clusters some theoretical calculations as well as experimental results [2] have shown rather a wide range of the values of $\epsilon$ between 0.3 to 2.

In case of bulk magnetite, Eq.(7) was verified for $\epsilon = 2$ with $M_s(0) = 502.5 \times 10^3$ A/m and B = $5.54 \times 10^{-7}$ K$^{-2}$, where the values of spontaneous magnetization were determined from the magnetization curves at different temperature [33]. A tentative fit of the temperature dependence of hyperfine field is shown in Fig. 9 using Eq. (7) for both $\epsilon = 2$ and 3/2. The fitting of $H_{hyp}(T)$ is found satisfactory for both the exponents within the experimental errors. In order to compare these results with those for bulk magnetite, the value of $H_{hyp}(0)$ is obtained to be 51.95 T with B = $21.54 \times 10^{-7}$ K$^{-2}$ for $\epsilon = 2$, indicating that the value of $J$ is reduced to ~0.5 of the bulk value.

## 4. Discussions

Temperature dependence of Mössbauer results indicates the evidence of typical features of superparamagnetic behaviour above 192 K for N10. In accordance with the Mössbauer results the magnetization curves well above $T_M$ do not show any remanance and coercivity for P5, P7, and N10, suggesting the characteristic features of superparamagnetism. In case of noninteracting nanoparticles, the superparamagnetic blocking temperature decreases with the particle size [1,2]. The blocking temperature is usually determined at the maximum in the temperature dependence of magnetization, where thermal energy becomes comparable to the anisotropy barrier [26,27]. In the present observation, the maximum in the temperature dependent magnetization does not show any consistency with the particle size, which indicates that $T_M$ is not a typical superparamagnetic blocking temperature of non-interacting particles, rather indicating non-negligible interparticle interactions. The interparticle interactions are mainly, (i) dipole-dipole interaction, (ii) exchange interaction through the surface of the particle. In case of polymer quoted particles the second term may be neglected, where the interparticle interaction is mainly dominated by the dipolar interaction. The anisotropic dipolar interaction favor ferromagnetic or antiferromagnetic alignments of the moments depending on the geometry, which may give rise to the necessary ingredients for the spin-glass states, namely, random distribution of easy axes associated with the magnetic frustration [1,23]. In addition, the surface effects for fine particles are also non-negligible in most of the cases because of the considerable increase of the surface spins to the total number of spins. Surface effect essentially results from the lack of translational symmetry at the boundaries of the particle because of lower coordination number and the existence of broken magnetic exchange bonds, which are responsible for the spin disorder or random spin canting associated with the occurrence of spin frustration [1]. In the present observation, the field cooled effect of magnetization does not indicate the typical feature of superparamagnetic behaviour, where the FC magnetization usually exhibits the increasing trend with the decreasing temperature below the superparamagnetic blocking temperature. In stead, the FC magnetization in the present observation, rather, exhibits a tendency of broad maximum below $T_M$, suggesting the feature of glassy behaviour.

If we look on the PAA quoted samples only, the values of $M_s$ decrease with the particle size (Table 2). The reduction of $M_s$ is commonly noticed for ferrimagnetic and antiferromagnetic oxides [1,34], in contrast to the enhancement of magnetization for few metallic ferromagnetic nanoparticles [35,36]. However, the reduction of magnetization in the oxide nanoparticles is a specific phenomenon of these materials, which is mainly due to the formation of a surface shell with spin disorder because of the competing antiferromagnetic interactions. This fact is interpreted by postulating the existence of a dead magnetic layer originated by the demagnetization of the surface spins, which causes a reduction of $M_s$ because of its paramagnetic behaviour [1,34]. The other possibility is the existence of random canting of the surface spins caused by the competing antiferromagnetic interactions between sublattices, which was proposed by Coey to account the reduction of $M_s$ in ferrimagnetic $\gamma$-$Fe_2O_3$ particles [37]. He found that even a magnetic field of 5 T was not enough to align all the spins along the field direction for particles with 6 nm in size. It has been commonly recognized that the decrease of $M_s$ indicates the misalignment of spins, though the origin of lack of full alignment of spins in the fine particles of ferrimagnetic oxides is still a subject of research, where no clear conclusion has not been established. In the present case, the origin of misalignment of spins is also not clear from the present experimental results. However, the nature of magnetization curve at 5 K with saturating tendency around 5 T does not fit satisfactorily with idea of spin canting, rather, suggests the possibility of magnetic dead layer on the particle surface.

## V. SUMMARY

We have demonstrated the magnetization and Mössbauer results on the polymer quoted magnetite nanoparticle with average particle size from 5.1 to 14.7 nm. The maximum in the temperature dependence of magnetization is found to be inconsistent with the particle size, where the maximum does not correspond to the blocking temperature. The temperature dependence of magnetization associated with the nature of field cooled effect indicates the existence of non-negligible dipolar interaction. The effective magnetic anisotropy energy densities increase with the decrease of particle size, which are suggested due to the increase of surface anisotropy. In accordance with the increase of anisotropy, the values of coercive field increase, while the decrease of remanence of magnetization with the increase of particle size is noticed at 5 K. On the other hand, the absence of coercivity and remanence of magnetization is observed well above $T_M$ indicating the characteristics of superparamagnetic behaviour. In accordance with the magnetic results the features of superparamagnetic relaxation are observed in the temperature dependent Mössbauer results. The temperature dependence of hyperfine field follows the similar dependence of saturation magnetization though the exchange integral is weakened to half of the bulk counterpart.

## ACKNOWLEDGEMENTS

The authors would like to thank Professor T. Kohara, Professor H. Kobayashi, and Professor H. Nakamura for the valuable discussions. This work is supported by CSIR, India. One of the authors (S. G.) wishes to thank JSPS, Japan for the support of Fellowship for Foreign Researcher. The magnetization data using SQUID magnetometer and Mössbauer spectra were measured in the University of Hyogo, Japan.

References

[1] Dormann J L, Fiorani D and Tronc E 1997 *Adv. Chem. Phys.* **XCVIII** 283

[2] Kodama R H 1999 *J. Magn. Magn. Mater.* **200** 359

[3] Luo W, Nagel S R, Rosenbaum T F and Rosensweig R E 1991 *Phys. Rev. Lett.* **67** 2721

[4] For example, Jonsson P E, Felton S, Svedlindh P and Norblad P 2001 *Phys. Rev.* B **64** 212402

[5] Zheng R K, Wen G H, Fung K K and Zhang X X 2004 *Phys. Rev.* B **69** 214431

[6] Frandsen C and Morup S 2005 *Phys. Rev. Lett.* **94** 027202

[7] Gruyters M 2005 *Phys. Rev. Lett.* **95** 077204

[8] Skumryev V, Stoyanov S, Zhang Y, Hadjipanayis G, Givord D and Nogues J 2003 *Nature* **423** 850

[9] D. C. Mattis 1988 *The Theory of Magnetism I* (Berline: Springer)

[10] Bragg W H 1915 *Phil. Mag.* **30** 305

[11] Neel L 1948 *Ann. Phys. Fr.* **3** 137

[12] Weiss P and Forrer R 1929 *Ann. Phys.* **12** 12279

[13] Rakhecha V C and Satya Murthy N S 1978 *J. Phys. C: Solid State Physics* **11** 4389

[14] Verwey E J W 1939 *Nature* **144** 327

[15] Garcia J and Subias G 2004 *J. Phys.: Condens Matter* **16** R145

[16] For example, Signorini L, Pasquini L, Savini L, Carboni R, Boscherini F, Bonetti E, Giglia A, Pedio M, Magne N and Nannarone S 2003 *Phys. Rev.* B **68** 195423

[17] Lin C R, Wang J S, Sung T W and Chiang R K 2005 *IEEE Tans. Mag.* **41** 3466

[18] Si S, Kotal A, Mandal T K, Giri S, Nakamura H and Kohara T 2004 *Chem. Mater.* **16** 3489

[19] El-Hilo M, O'Grady K and Chantrell R W 1992 *J. Magn. Magn. Mater.* **114** 295

[20] Blanco-Mantecon M and O'Grady K 2006 *J. Magn. Magn. Mater.* **296** 124

[21] Jonsson T, Mattsson J, Djurberg C, Khan F A, Norblad P and Svedlindh P 1995 *Phys. Rev. Lett.* **75** 4138

[22] Akulov N 1931 *Z. Physics.* **67**, 194; Gans R 1932 *Ann. Phys. (Leipzig)* **15** 28

[23] Batle X and Labarta A 2002 *J. Phys. D: Appl. Phys.* **35** R15

[24] R. C. O'Handley, *Modern Magnetic Materials* (Wiley, New York, 2000).

[25] Kronmuller H, Fischer R, Bachmann M and Leineweber T 1999 *J. Magn. Magn. Mater.* **203** 12

[26] Neel L 1949 *Ann. Geophys.* **5** 99

[27] Brown J W F 1963 *Phys. Rev.* **130** 1677

[28] Tronc E, Prene P, Jolivet J P, D'Orazio F, Lucari F, Fiorani D, Godinho M, Cherkaui R, Nogues M, Dormann J L 1995 *Hyperfine Interact.* **95** 129

[29] For example, Van. Lierop J and Ryan D H 2000 *Phys. Rev. Lett.* **85** 3021

[30] Brand R A 1987 *Nuclear Instr. Methods* B **28** 398

[31] Berry F J, Skinner S and Thomas M F 1998 *J. Phys.: Condens. Mater* **10**, 215

[32] Hendriksen P V, Linderoth S, Lindgard P – A 1993 *Phys. Rev.* B **48** 7259

[33] Dillon J. F 1962 *Landolt-Börnstein New Series Group III* (Berlin: Springer), vol 2(9), ch 29 p. 50-1

[34] Berkowitz A E, Shuele W J and Flanders P J 1968 *J. Appl. Phys.* **39** 1261

[35] Billas L M L, Chatelain A and de Herr W A 1994 *Science* **265** 1682

[36] Apsel S E, Emmert J W, Deng J and Bloomfield L A 1996 *Phys. Rev. Lett.* **76**, 1441

[37] Coey J M D 1971 *Phys. Rev. Lett.* **27** 1140

Table 1: The characteristic features of magnetite nanoparticles

| Sample | Poly | W (%) | $D_{TEM}$ (nm) | $\sigma$ |
|--------|------|-------|----------------|----------|
| P5     | PAA  | 13.1[a] | 5.1 | 0.42 |
| P7     | PAA  | 12.1[a] | 7.0 | 0.58 |
| P15    | PAA  | 11.7[a] | 14.7 | 4.20 |
| N10    | NaCMC | 13.4[a] | 10.5 | 0.60 |

Ploy: polyelectrolyte. W(%): percentage of weight of polyelectrolyte in the powdered samples. $\sigma$: standard deviation, $\sigma = [\Sigma n_i(D_i - D_{TEM})^2/(N-1)]^{0.5}$, $n_i$ being the number of particles having diameter $D_i$, $D_{TEM}$ being the average diameter $= (\Sigma n_i D_i)/N$, $N$ being the total number of particles. [a]Error = ± 0.5

Table 2: Effective anisotropy energy density, $K_{an}$, upper limit of magnetic size, $D_M$, magnetization at 5 T above $T_M$, $M_s(T)$, coercive field, $H_c$, remanance magnetization, $M_r$, magnetization at 5 T and 5 K, $M_s$.

| | From magnetization curves above $T_M$ (Figs. 2 & 3) | | | From magnetization curves at 5 K (Fig 4) | | |
|---|---|---|---|---|---|---|
| Sample | $D_M$ (nm) | $M_s(T)$ ($\mu_B$) | $K_{an}$ ($10^5$ J/m$^3$) | $H_c$ (T) | $M_r$ ($\mu_B$) | $M_s$ ($\mu_B$) |
| P5  | 8.0[a]  | 1.42[b] | 0.18[c] | 0.0345[d] | 0.48[b] | 1.79[b] |
| P7  | 9.0[a]  | 1.40[b] | 0.12[c] | 0.0267[d] | 0.53[b] | 1.82[b] |
| P15 | -       | -       | -       | 0.0232[d] | 0.62[b] | 1.88[b] |
| N10 | 10.5[a] | 1.21[b] | 0.09[c] | 0.0260[d] | 0.61[b] | 1.75[b] |

[a]Error = ± 0.25
[b]Error = ± 0.02
[c]Error = ± 0.002
[d]Error = ± 0.0005

Table 3: Fitted Mössbauer parameters at different temperatures.

| T (K) | IS (mm/s) | QS (mm/s) |
|---|---|---|
| 4.2 | 0.39[a] | -0.02[b] |
| 21 | 0.45[a] | -0.02[b] |
| 77 | 0.44[a] | -0.03[b] |
| 102.5 | 0.42[a] | -0.03[b] |
| 123.5 | 0.40[a] | -0.03[b] |
| 170 | 0.39[a] | -0.03[b] |
| 192 | 0.40[a] | -0.03[b] |

IS: isomer shift. QS: quadrupole splitting. [a]error = ± 0.01. [b]error = ± 0.005

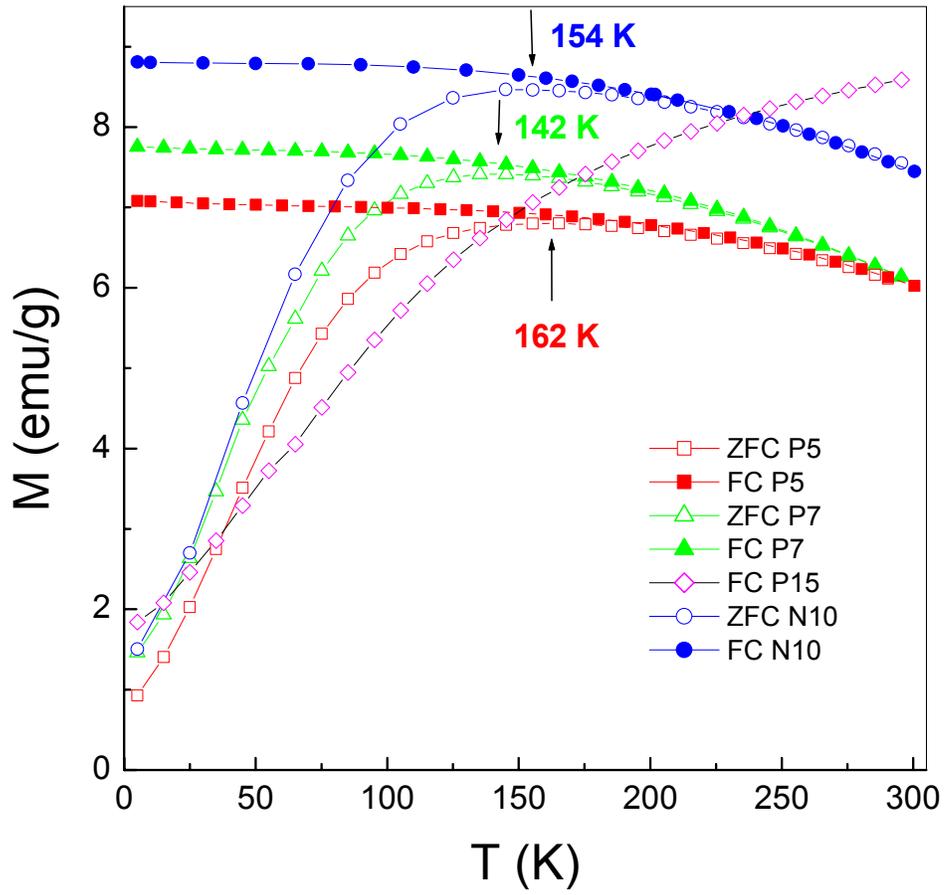

FIG. 1

FIG. 1. Temperature dependence of zero-field cooled (ZFC) magnetization for P5, P7, P15, and N10 and field cooled (FC) magnetization for P5, P7, and N10. The arrows indicate the maximum in the temperature dependent ZFC magnetization.

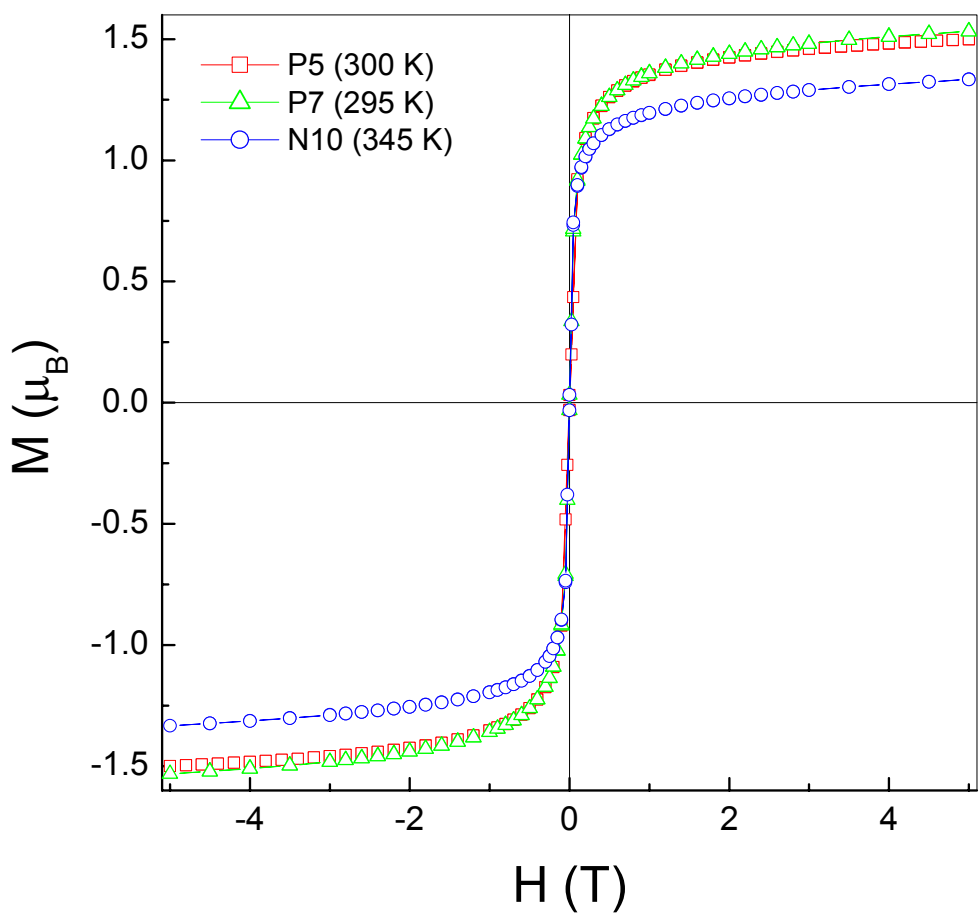

FIG. 2

FIG. 2. Magnetization curves for P5, P7, and N10 at 300 K, 295 K, and 345 K, respectively.

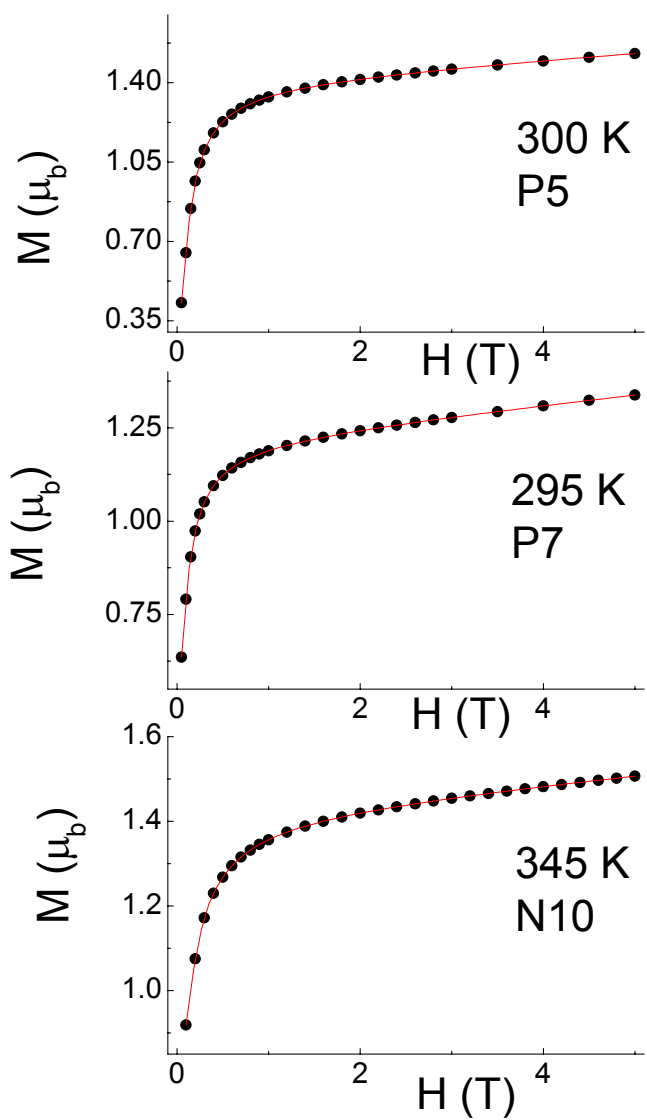

FIG. 3

FIG. 3. Fitting of the magnetization curves of P5, P7, and N10 using Eq. (3) described in the text.

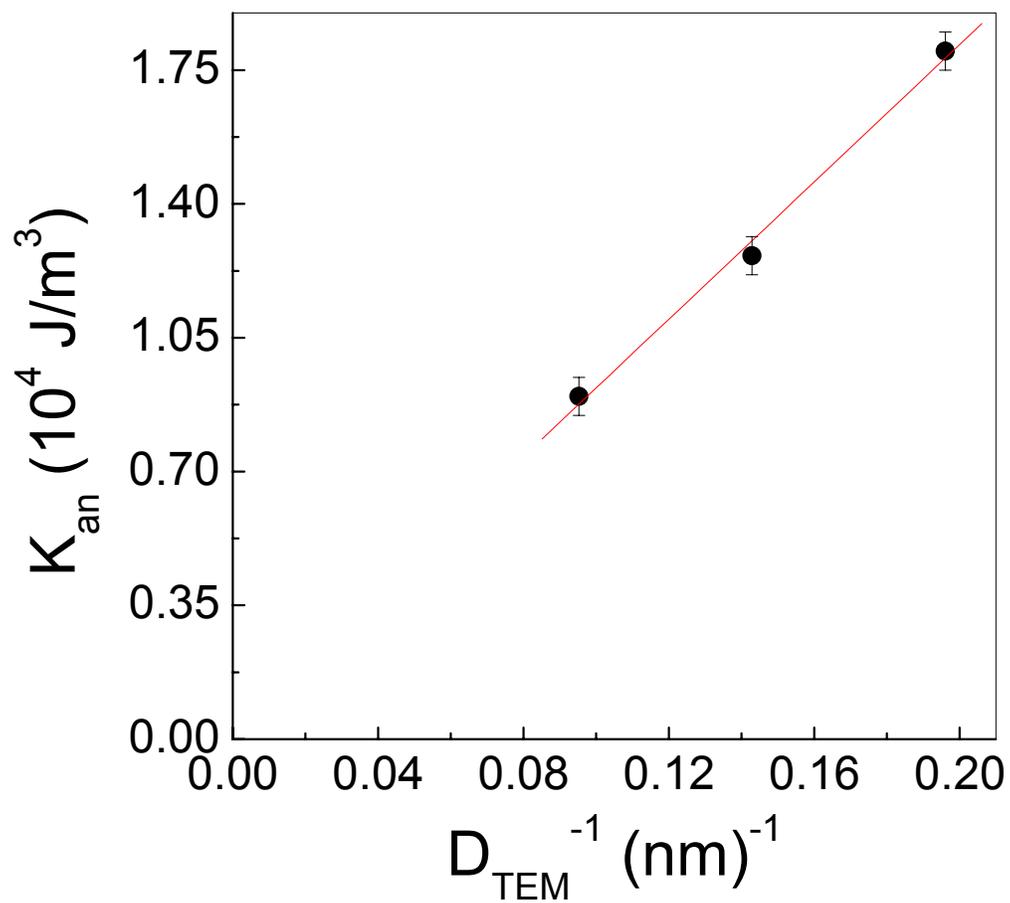

FIG. 4

FIG. 4. Plot of effective anisotropy ($K_{an}$) against inverse of the particle size ($D_{TEM}$) obtained from Transmission Electron Microscope.

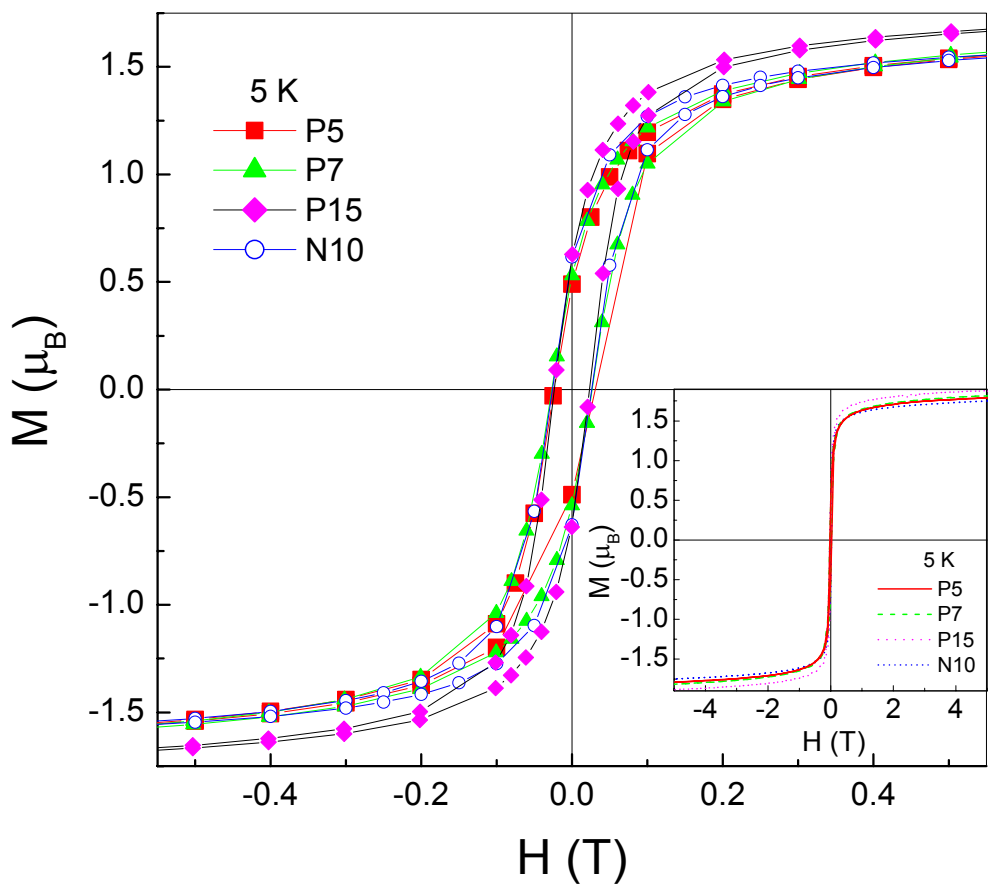

FIG. 5

FIG. 5. Hysteresis of magnetization at 5 K for P5, P7, P15, and N10 in the low field range indicating the nature remanance and coercivity. The inset exhibits the examples of the same up to 5 T.

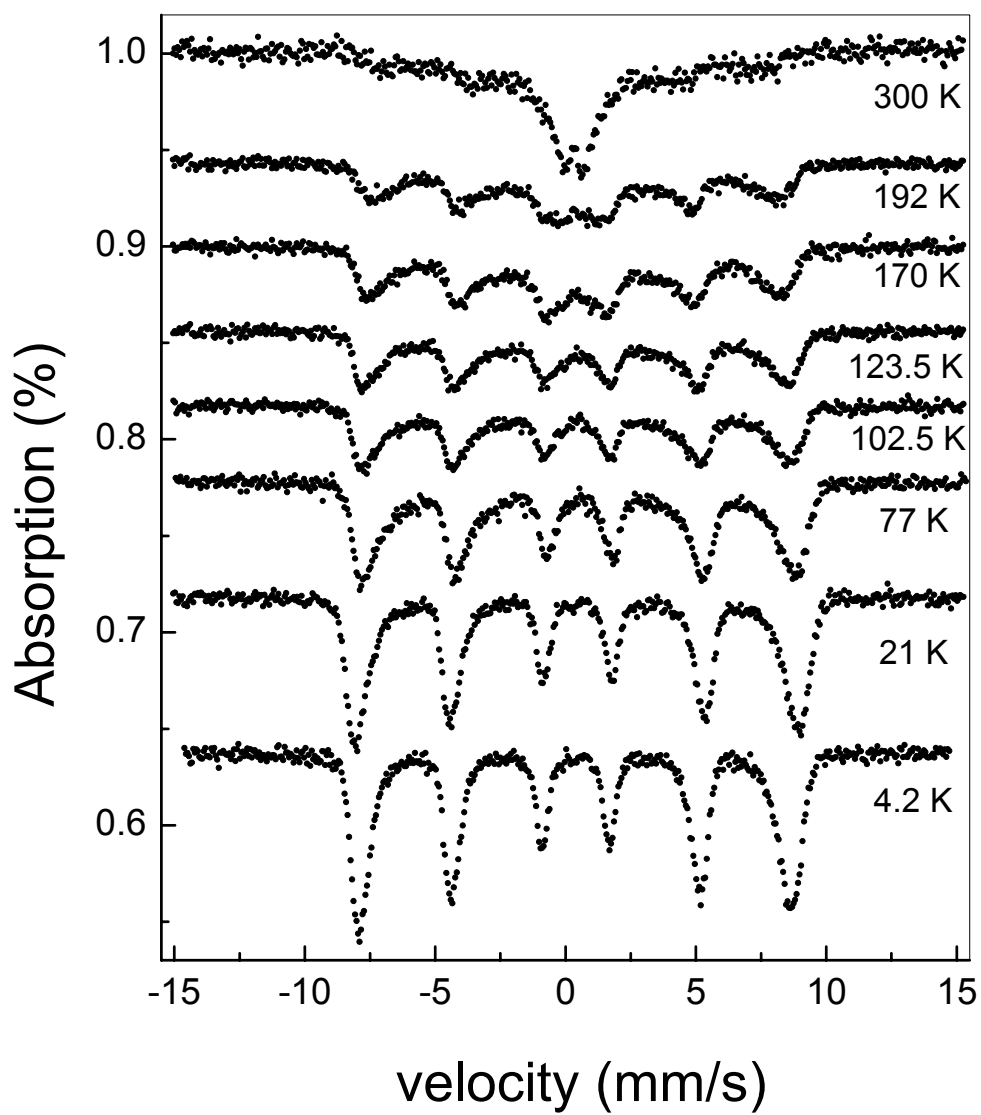

FIG. 6. The characteristic features of Mössbauer spectra at different temperatures for N10.

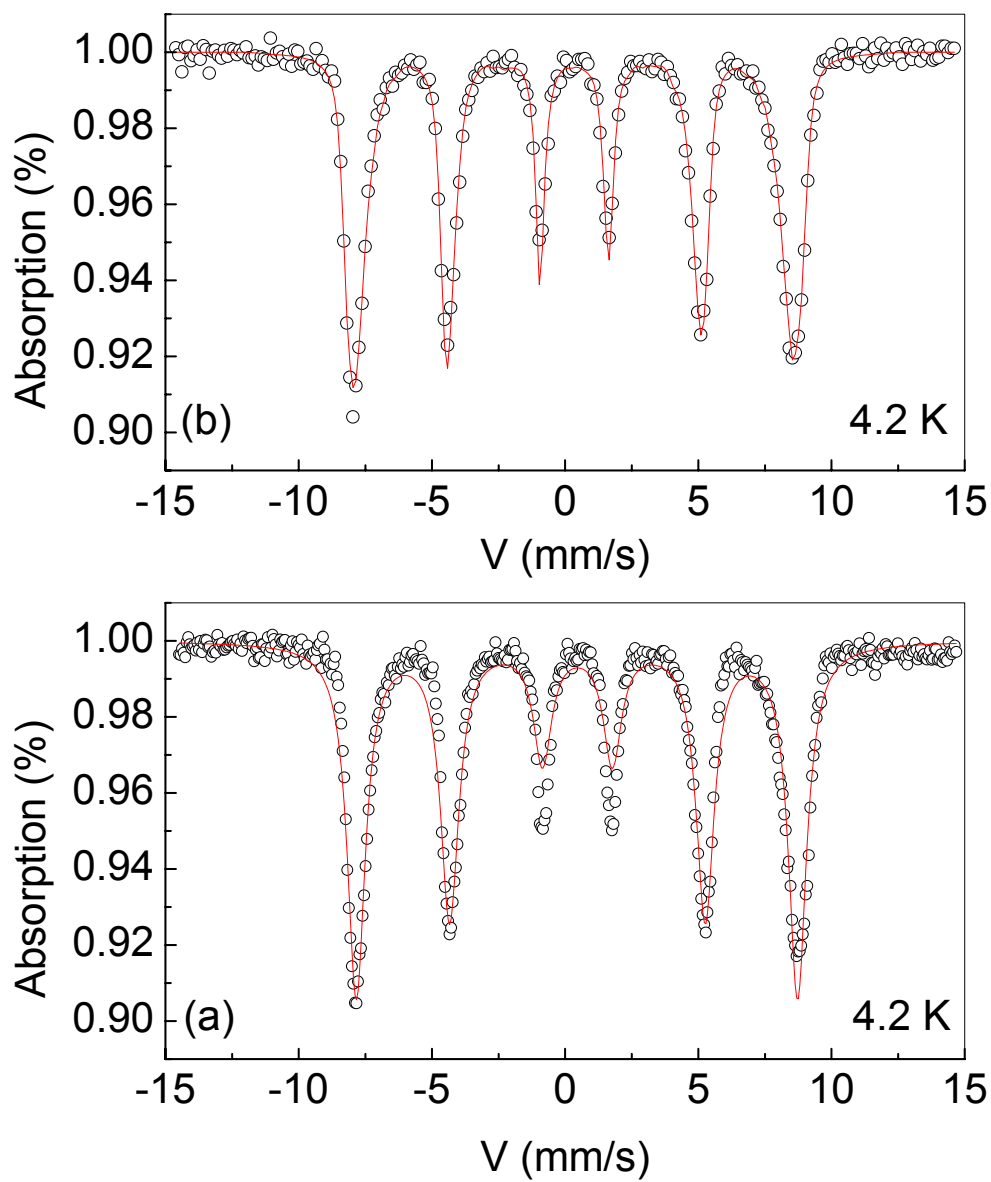

FIG. 7

FIG. 7. Fitting of Mössbauer spectrum at 4.2 K considering unique hyperfine field (a) and a distribution of hyperfine field (b) for N10.

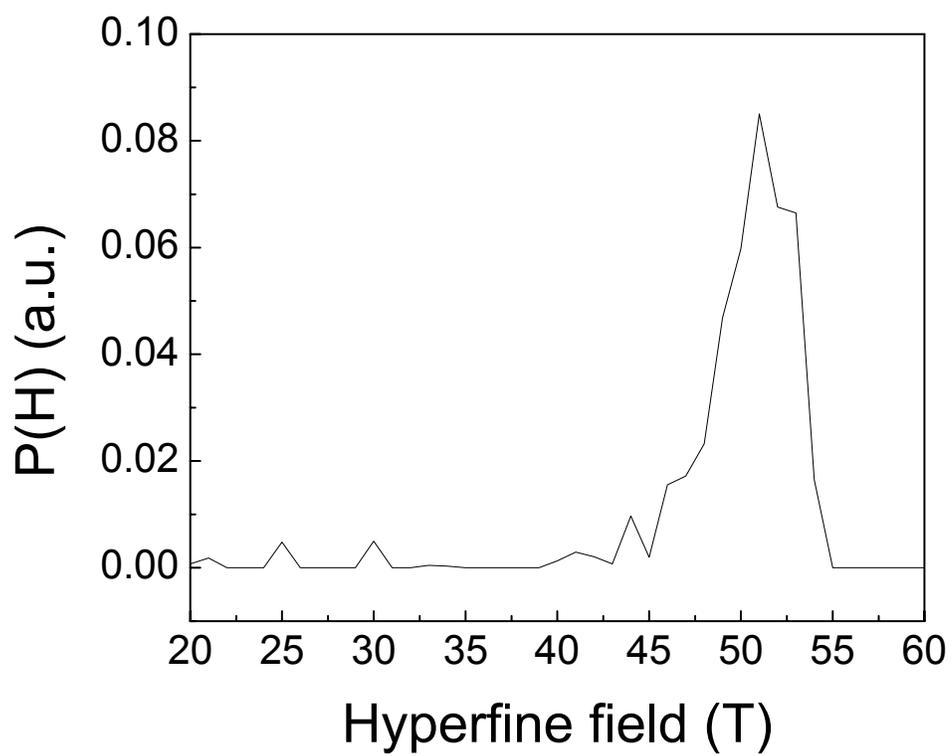

FIG. 8

FIG. 8. Distribution function, *P*(*H*) in arbitrary unit against the hyperfine field for N10.

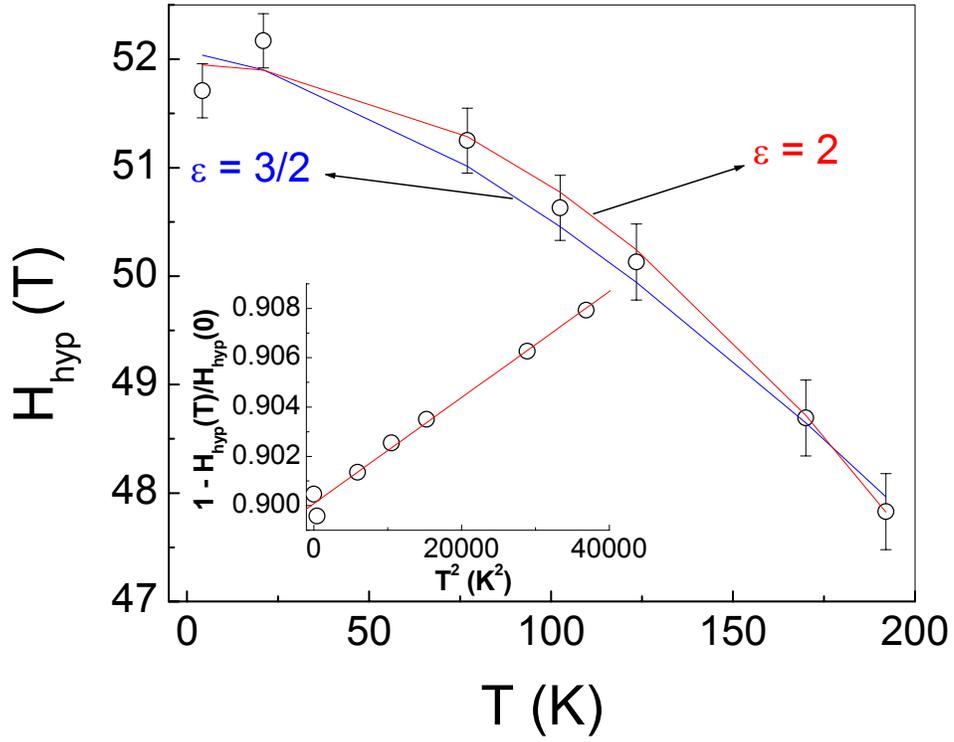

FIG. 9

FIG. 9. Temperature dependence of the hyperfine field. The red curve indicates the fitting using Eq (7) in the text with $\varepsilon = 2$ and the blue curve indicates the same with $\varepsilon = 3/2$. The inset shows the plot of $1 - H_{hyp}(T)/H_{hyp}(T)$ against $T^2$ and the red continuous line shows the fitting with $\varepsilon = 2$.